\documentclass[epsfig,twocolumn]{IEEEtran}
\usepackage{subfigure,times,epsfig,epsf,latexsym,graphicx,multirow}
\usepackage{stfloats,amsmath,enumerate}
\usepackage{algorithm,algpseudocode}
\usepackage{cite,url}

\newtheorem{example}{Example}

\begin{document}



\title{A New Energy Efficient MAC Protocol based \\ on Redundant
Radix for Wireless Networks}

\author{
Koushik~Sinha \\
Honeywell Technology Solutions, Bangalore, India \\
Email: sinha\_kou@yahoo.com \\
}

\date{}
\maketitle

\begin{abstract}
In this paper, we first propose a redundant radix based number (RBN)
representation for encoding the data to be transmitted in a wireless
network. This RBN encoding uses three possible values - 0, 1 and
$\bar 1$, for each digit to be transmitted. We then propose to use
silent periods (zero energy transmission) for transmitting the 0's
in the RBN encoded data thus obtained. This is in contrast to most
conventional communication strategies that utilize energy based
transmission (EbT) schemes, where energy expenditure occurs for
transmitting both 0 and 1 bit values. The binary to RBN conversion
algorithm presented here offers a significant reduction in the
number of non-zero bits in the resulting RBN encoded data. As a
result, it provides a highly energy-efficient technique for data
transmission with silent periods for transmitting 0's. We simulated
our proposed technique with ideal radio device characteristics and
also with parameters of various commercially available radio
devices. Experimental results on various benchmark suites show that
with ideal as well as some commercial device characteristics, our
proposed transmission scheme requires 69\% less energy on an
average, compared to the energy based transmission schemes. This
makes it very attractive for application scenarios where the devices
are highly energy constrained. Finally, based on this transmission
strategy, we have designed a MAC protocol that would support the
communication of such RBN encoded data frames.
\end{abstract}

\section{Introduction} \label{intro}

During recent years, wireless ad hoc networks have received
considerable attention of researchers for their increasing
applications in various fields, e.g, military communications,
disaster relief, rescue operations, etc. There exist different
schemes for transmitting data in a wireless network. Depending on
the situation, either both 0 and 1 are represented by non-zero
voltage levels, or one of the bit values is represented by a zero
voltage level while a non-zero voltage level is used to distinguish
the other bit value. An example of the latter is the polar
return-to-zero (polar-RZ) transmission scheme, where a 0 corresponds
to a zero voltage level, while a 1 is represented by a non-zero
voltage level. However, most existing transmission schemes utilize
non-zero voltage levels for both 0 and 1 so as to distinguish
between a silent and a busy channel. Communication strategies that
require energy expenditure for transmitting both 0 and 1 bit values
are known as {\em energy based transmission} (EbT) schemes. For
example, in order to communicate a value of 278, a node will
transmit the bit sequence $<1, 0, 0, 0, 1, 0, 1, 1, 0>$, consuming
energy for every bit it transmits. Thus, if the energy required per
bit transmitted is $e_b$, the total energy consumed to transmit the
value 278 would be $9e_b$.

In this paper, we propose a communication technique that first
recodes a binary data in redundant radix based number (RBN)
representation \cite{tagaki85} and then uses silent periods to
communicate the bit value of '0'. We show that by using the
redundant binary number system (RBNS) that utilizes the digits from
the set \{-1, 0, 1\} to represent a number with radix 2, we can
significantly reduce the number of non-zero digits that need to be
transmitted. The transmission time remains linear in the number of
bits used for data representation, as in the binary number system.
We finally propose a MAC protocol that would support the
communication of such RBN encoded data frames with a significant
amount of energy savings.

We have simulated our proposed transmission algorithm with both
ideal device characteristics and parameters of several commercially
available radio devices. The results of these experiments show that,
for ideal device characteristics and even for some commercial device
characteristics, the increase in energy savings with our proposed
algorithm over the existing energy based transmission schemes is, on
an average, equal to 69\%.

\section{Related Work} \label{related}

Recent research efforts on reducing energy consumption have mainly
been focussed on the MAC layer design \cite{demirkol05}, optimizing
data transmissions by reducing collisions and retransmissions
\cite{sinha04} and through intelligent selection of paths or special
architectures for sending data \cite{enz04}. A survey of routing
protocols in wireless sensor networks can be found in
\cite{demirkol05}. In all such schemes, the underlying communication
strategy of sending a string of binary bits is {\em energy based
transmissions} (EbT) \cite{zhu05,chen05}, which implies that the
communication of any information between two nodes involves the
expenditure of energy for the transmission of data bits. In
\cite{zhu05}, a new communication strategy called {\em Communication
through Silence} (CtS) has been proposed that involves the use of
silent periods as opposed to energy based transmissions. CtS,
however, suffers from the disadvantage of being exponential in time.
An alternative strategy, called Variable-base tacit communication
(VarBaTaC) has been proposed in \cite{chen05} that uses a variable
radix-based information coding coupled with CtS for communication.

\section{Preliminaries and Proposed Communication Scheme} \label{prelim}

The redundant binary number system (RBNS) \cite{tagaki85} utilizes
the digits from the set \{-1, 0, 1\} for representing numbers using
radix 2. In the rest of the paper, for convenience, we denote the
digit '-1' by $\bar{1}$. In RBNS, there can be more than one
possible representation of a given number. For example, the number 7
can be represented as either $111$ or $100\bar{1}$ in RBNS. In this
work, we utilize this property of RBNS to recode a message string so
as to reduce the number of 1's in the string while transmitting the
message \cite{phdthesis}. The original binary message can, however,
be obtained at the receiver end by reconverting the received message
from RBN to binary number system \cite{tagaki85}.

The basic idea of our recoding scheme is as follows : Consider a run
of $k$ 1's, $k > 1$. Let $i$ be the bit position for the first 1 in
this run, $i \geq 0$ (bit position 0 refers to the least significant
bit at the rightmost end). Let $v$ represent the value of this run
of $k$ 1's. Then,

\begin{equation} \label{rbneqn1}
v = 2^i + 2^{i + 1} + 2^{i + 2} + \ldots + 2^{k + i - 1}
\end{equation}

Alternatively, we can rewrite equation~\ref{rbneqn1} as,

\begin{equation} \label{equation2}
v = 2^{k + i} - 2^i
\end{equation}

Equation~\ref{equation2} can be represented in RBNS by a '1' at bit
position $(k + i)$ and a $\bar{1}$ at bit position $i$, while all
the intermediate 1's between them are converted to 0's. Thus, a long
run of 1's can equivalently be replaced by a run of 0's and only two
non-zero digits, 1 and $\bar{1}$.

Observe that for a run of $k$ 1's, $k > 1$, the savings in terms of
the number of non-zero digits is $k - 2$. However, the number of
non-zero digits remain unchanged for $k = 2$.

Thus, if we keep the transmitter switched-off for 0 bit-values, the
power consumption of the transmitter will be less than that in {\em
energy based transmission} (EbT) schemes. Hence, by combining this
approach of silent zero transmission with our RBNS-based recoding
strategy, a significant reduction in the energy expenditure during
data transmission can be achieved when compared to the energy based
transmission (EbT) of binary data.

Our proposed low energy transmission strategy involves the execution
of the following two steps : \\

\noindent {\bf Algorithm TransmitRBNData}


{\bf Step 1} : Recode the $n$-bit binary data frame to its
equivalent RBNS data frame using steps 1.1 and 1.2 stated below.


{\bf Step 1.1} : Starting from the least significant bit (lsb)
position, scan the string for a run 1's of length $> 1$. A run of
$k$ 1's ($k > 1$) starting from bit position $i$, is replaced by an
equivalent representation consisting of a '1' at bit position $k +
i$ and a $\bar{1}$ at bit position $i$, with 0's in all intermediate
bit positions. \label{reduction1}

{\bf Step 1.2} : Every occurrence of the bit pattern $\bar{1}1$ in a
string obtained after step 1.1, is replaced by the equivalent bit
pattern $0\bar{1}$. \label{reduction2}


{\bf Step 2} : Transmit the RBNS data frame obtained from step 1
above.


Note that the encoding process of an $n$-bit binary string to its
equivalent RBNS representation can result in a RBNS string of length
of either $n$ or $n + 1$ symbols. If a run of 1's of length $> 1$
ends in the most significant bit (msb), then by virtue of step 1.1
of {\em TransmitRBNData} algorithm, the symbol $1$ is placed at the
position $msb + 1$. Otherwise, if the msb was 0, then the RBNS
string also has exactly $n$ symbols.

\begin{example} \label{examplereduction}
Consider in a given binary string, a substring, say $110111$, with
only one '0' trapped between runs of 1's. Then following step 1.1,
we would get the string $10\bar{1}100\bar{1}$. Note the presence of
the pattern $\bar{1}1$ for this trapped '0'. Application of step 1.2
of algorithm TransmitRBNData  to the bit pattern $\bar{1}1$ replaces
it by $0\bar{1}$, thus resulting in a further reduction in the
number of non-zero symbols to be transmitted.
\end{example}

We now present the receiver side algorithm to receive a RBNS data
frame and convert it back to binary : \\

\noindent {\bf Algorithm ReceiveRBNData}


{\bf Step 1} : Receive the RBNS data frame in a buffer, say
$recv\_buf$.

{\bf Step 2} : Set $runflag \leftarrow false$. Now starting from
lsb, scan the RBNS string in $recv\_buf$ sequentially to obtain the
binary equivalent using steps 2.1 through 2.3 :


{\bf Step 2.1} : If the $i^{th}$ bit, $recv\_buf[i] = \bar{1}$ then
execute steps 2.1.1 and 2.1.2, otherwise execute step 2.2 :


{\bf Step 2.1.1} : If $runflag = false$ then the corresponding
output binary bit is 1. Also set $runflag \leftarrow true$.

{\bf Step 2.1.2} : If $runflag = true$ then the corresponding output
binary bit is 0.


{\bf Step 2.2} : If the $i^{th}$ bit, $recv\_buf[i] = 1$ then
execute steps 2.2.1 and 2.2.2, otherwise execute step 2.3 :


{\bf Step 2.2.1} : If $runflag = true$ then the corresponding output
binary bit is 0. Also set $runflag \leftarrow false$.

{\bf Step 2.2.2} : If $runflag = false$ then the corresponding
output binary bit is 1.


{\bf Step 2.3} : If the $i^{th}$ bit, $recv\_buf[i] = 0$ then
execute steps 2.3.1 and 2.3.2 :


{\bf Step 2.3.1} : If $runflag = true$ then the corresponding output
binary bit is 1.

{\bf Step 2.3.2} : If $runflag = false$ then the corresponding
output binary bit is 0.



{\bf Step 3} : Set $i \leftarrow i + 1$ and repeat from step 2 until
the entire received RBNS data frame is scanned and converted to the
binary equivalent. The equivalent binary data is then passed onto
the higher layers of the network stack.

\begin{table*}
\centering \caption{{\small Number of occurrences of runlengths for
$n = 8$}} \label{table1}
\begin{tabular}{|c|c|c|}

\hline

Runlength size ($k$) & Number of maximum & Number of \\
& \multicolumn{1}{c|}{possible values of $i_k$} &
\multicolumn{1}{c|}{occurrences} \\

\hline \hline

1 & 4 & 320 \\
2 & 3 & 144 \\
3 & 2 & 64 \\
4 & 1 & 28 \\
5 & 1 & 12 \\
6 & 1 & 5 \\
7 & 1 & 2 \\
8 & 1 & 1 \\

\hline

\end{tabular}
\end{table*}

We note that the application of steps 1.1 and 1.2 of the {\em
TransmitRBNData} algorithm ensures that the bit patterns $1\bar{1}$
and $\bar{1}1$ can not occur in the transmitted data. Hence, there
is only a unique way of converting the received RBNS data into its
binary equivalent.

\section{Analysis of the Energy Savings} \label{analysis}

We denote a run of 1's of length $k$ by $R_k$. Let us append a zero
on left of each such $R_k, 1 \leq k \leq n$ and denote the symbol $0
\ R_k$ by $y_k$. We also denote a  single zero by the symbol $y_0$.
Then each such $y_k, 0 \leq k \leq n$, will be a string of length
$k+1$. To find out the total number of occurrences of $R_k$'s, $1
\leq k \leq n$, in all possible $2^n$ strings of length $n$, we
would first compute the total number of occurrences of exactly $i_k$
number of $y_k$'s. Let this number be denoted by the symbol
$N_n^{i_k, k}$.

We use a generating function based approach to derive an expression
for $N_n^{i_k, k}$ in all possible binary strings of length $n$. The
detailed analysis is given in \cite{phdthesis}. We have omitted it
here for the sake of brevity and state only the final result as
follows :

For a given $n$ and $k \geq 1$, $N_n^{i_k, k}$ is given by,

\begin{eqnarray*}
N_n^{i_k, k} &= &i_k \sum_{r=1}^{n+1-(k+1)i_k} {{r + i_k} \choose {i_k}} \sum_{q=0}^r \sum_{j=0}^{q} (-1)^{q+j} \times \\
& &{{r+m-1-kq-j} \choose {m-kq-j}} {r \choose q} {q \choose j} \\
\end{eqnarray*}

\begin{example} \label{example1}
For $n=8$, $k = 2$ and $i_k = 2$, we get the number :

\begin{eqnarray*}
N_8^{2,2} & = & 2 {3 \choose 2} \biggl[{2 \choose 2} {1 \choose 0} {0 \choose 0} \\
   & &- {0 \choose 0} {1 \choose 1} {1 \choose 0} + {{-1} \choose {-1}} {1
\choose 1} {1 \choose 1}\biggr] \\
& &+ 2 {4 \choose 2} \biggl[{2 \choose 1} {2 \choose 0} {0 \choose
0} - 0 + 0 \biggr] \\
& & + 2 {5 \choose 2} \biggl[{2 \choose 0} {3 \choose 0} {0 \choose 0} - 0 + 0 \biggr] \\
  & = & 2(0 + 12 + 10) = 44\\
\end{eqnarray*}
\end{example}

For a given $k \geq 1$, if we now sum the expression $N_n^{i_k, k}$
for all possible values of $i_k$, $1 \leq i_k \leq \lfloor
(n+1)/(k+1) \rfloor$, then we get the total number of occurrences of
$R_k$ in all possible strings of length $n$. Table~\ref{table1}
shows the total number of occurrences of all possible runlengths of
1's in all possible binary strings of length 8 bits. It was shown in
\cite{phdthesis} that considering all possible $2^n$ binary strings
of length $n$ each, the total number of 1's and $\bar{1}$'s in the
RBN coded message after applying both the steps 1.1 and 1.2 of the
algorithm TransmitRBNData would be $(n + 2)2^{n-2}$.

\section{Experimental Results} \label{results}

\begin{figure*}
\centerline{\psfig{file=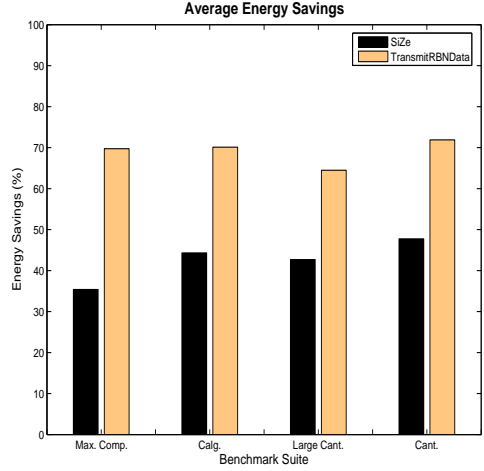,width=2.5in,height=2.5in}}
\centering \caption{Comparison of Average Energy Savings for the
Benchmark Test Suites} \label{figure2}
\end{figure*}

Experimental results demonstrate that algorithm {\em
TransmitRBNData} significantly reduces the energy consumption
required for transmission, for different types of application
scenarios. We tested our algorithm on several popular compression
benchmark test suites \cite{maxcomp,canterbury}. The results for
these test suites are presented in figure~\ref{figure2}. We have
omitted the detailed results for each file of the individual
benchmark suites for the sake of brevity. For the purpose of the
experiments, we assumed a data frame size of 1024 bits. All the
reported values in figure~\ref{figure2} are with respect to energy
based transmission schemes where the transmission of both '0' and
'1' bit values require the expenditure of energy. The column "SiZe"
mentioned in the tables and figures, refers to a silent zero (SiZe)
transmission scheme introduced in \cite{phdthesis}.

\begin{table*}
\centering \caption{{\small Theoretical energy savings results for
different radios, $n = 1024$}} \label{tabledevices}
\begin{tabular}{|l||c|c|c|c|}
\hline Vendor & Maxim & Chipcon & RFM & Maxim \\
part no. & 2820 & CC2510Fx & TR1000 & 1479 \\
\hline \hline

Data rate (kbps) & 50 & 2.5 & 25 & 2.0 \\
Symbol duration ($\mu s$) & 20 & 400 & 40 & 500 \\
$V_{cc}$ (volts) & 2.7 & 3.0 & 3.0 & 2.7  \\
TX state, $I_{high}$ (mA) & 70.0 & 23.0 & 12.0 & 7.3 \\
Active state, $I_{low}$ (mA) & 25.0 & 7.5 & $7.0 \texttt{x} 10^{-4}$  & $0.2 \texttt{x} 10^{-6}$  \\
$t_{on}$ ($\mu s$) & 3 & 195 & 16 & 200 \\
\hline

$\gamma_{SiZe}$ & 32.14\% & 33.69\%
& 50.0\% & 50.0\% \\

$\gamma_{dev}$ & 48.18\% &
50.51\% & 74.95\% & 74.95\% \\

\hline
\end{tabular}

\end{table*}

Considering the mean of the values reported for the {\em SiZe
protocol}, we find that binary encoded files consists of 42.5\%
zeroes on an average which thus translates into an increased energy
savings of 42.5\%, as compared to an EbT transmission scheme.
Application of algorithm TransmitRBNdata on binary encoded files to
create RBN encoded files cause on an average, an increased savings
in energy from 42.5\% to 69\%, when averaged over the values
reported in figure~\ref{figure2}.

Experimental results also showed that increasing the data frame size
increases the fractional savings in energy as longer runs of ones
can then be reduced. It increases steeply with the increase in frame
size, when the size of the frames is small (8, 16, 32, 64, $\ldots$
bits) and plateaus out for larger frame sizes. We observed that in
general, for frame sizes larger than 1024 bits, the increase in
fractional savings is either very small or none.

The results show that the maximum increase in energy savings
(34.4\%) with our proposed algorithm over the SiZe protocol is
obtained for the Maximum Compression test suite \cite{maxcomp},
while the minimum (21.8\%) is for the Large Canterbury suite
\cite{canterbury}. From the results in figure~\ref{figure2} we see
that there is an increase of $ 69\% - 42.5\% = 26.5\%$ in
transmission energy savings, when averaged over all benchmark suites
considered in the figure, by using the proposed {\em
TransmitRBNData} algorithm over the SiZe protocol.

\subsection{Results Considering Device Characteristics}

\begin{figure*}
\centerline{\hbox{ \hspace{0.0in}
    \subfigure{\psfig{file=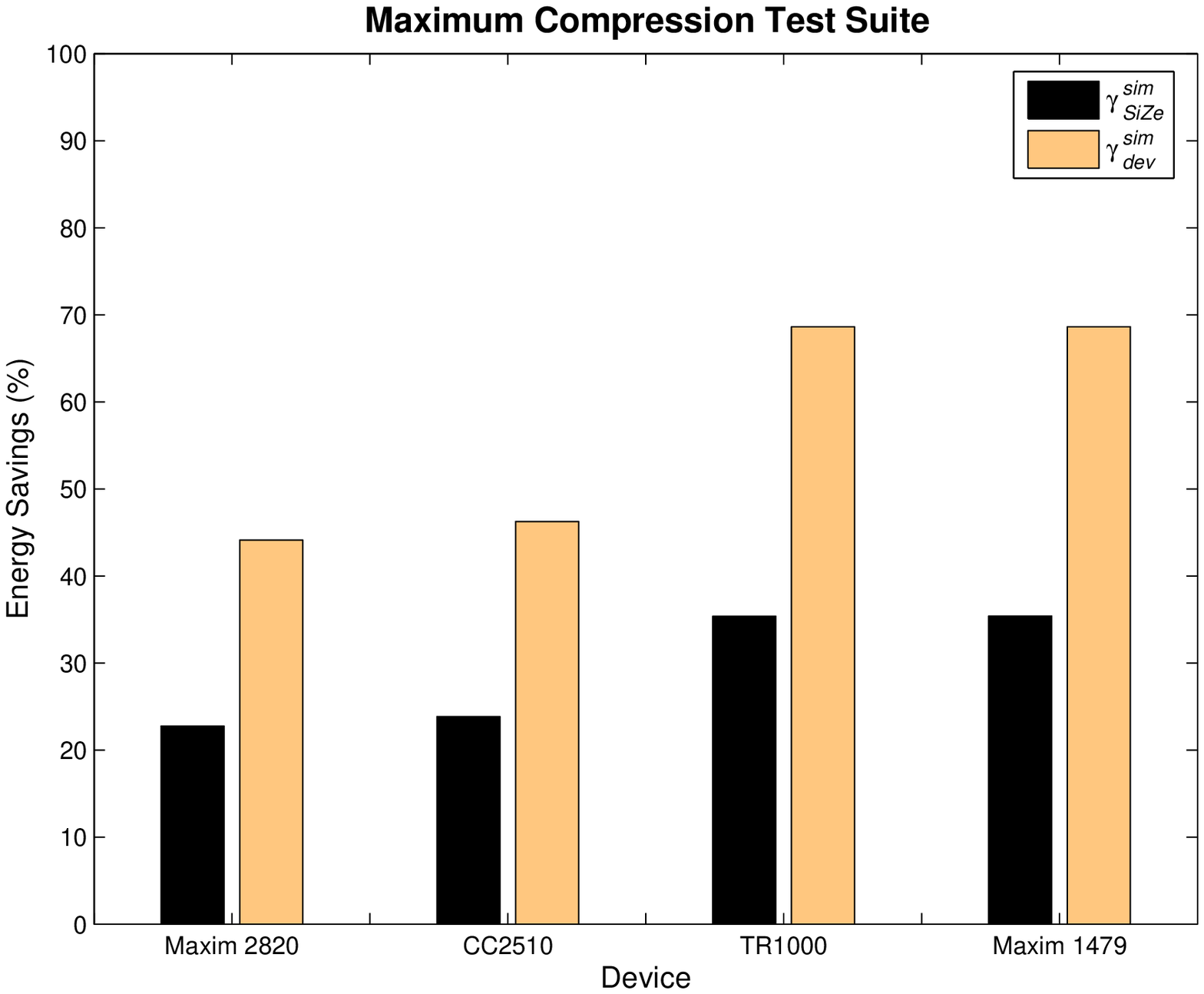,width=2.5in,height=2.5in}
    \hspace{0.13in}
    \label{figure3:a}}
    \subfigure{\psfig{file=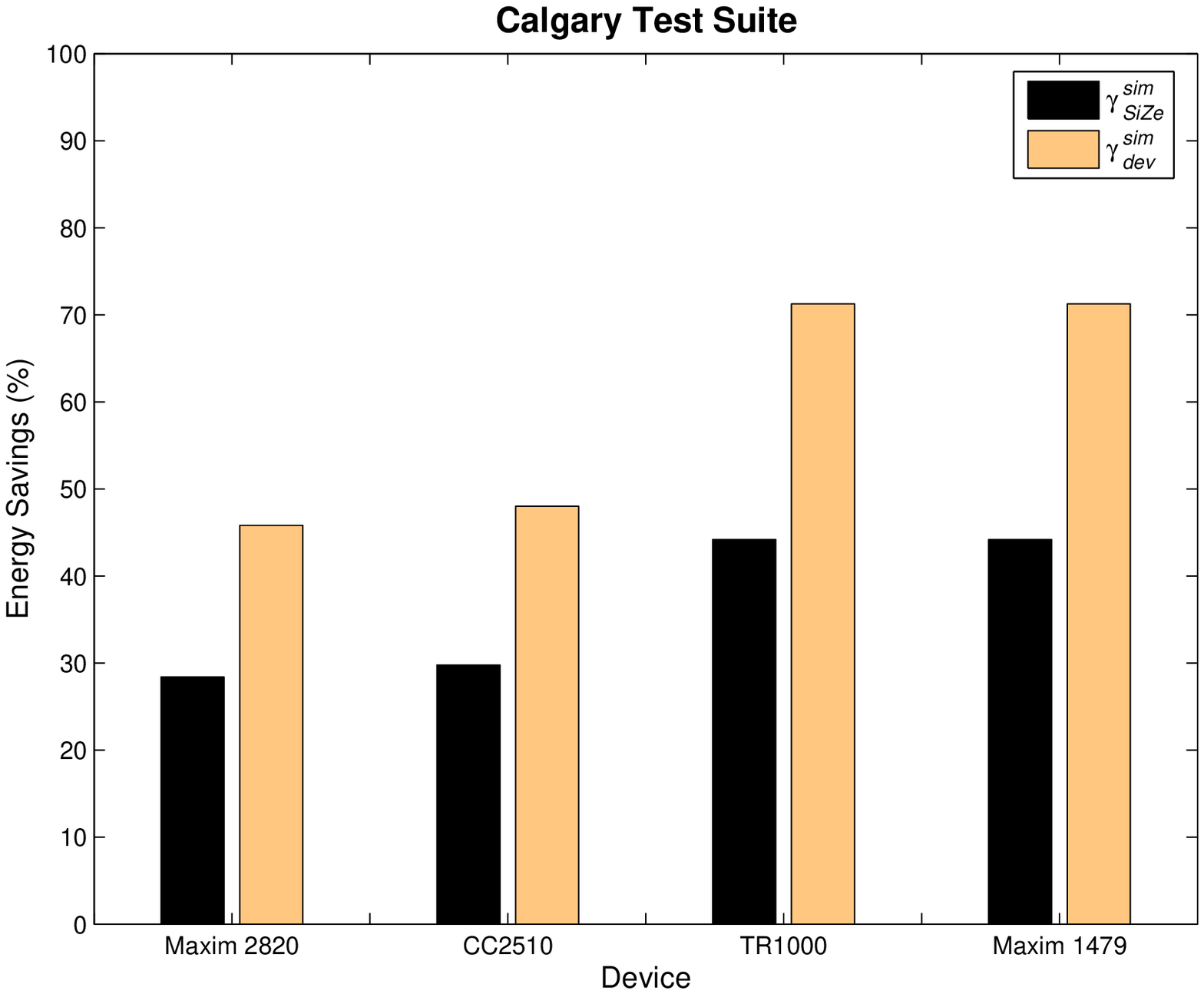,width=2.5in,height=2.5in}
    \label{figure3:b}}
    }
  }

  \hbox{\hspace{1.0in} (a) \hspace{2.67in} (b)}
  \caption{\small{Device Specific Energy Savings for the Maximum
Compression and Calgary Test Suites}}
  \label{figure3}
\end{figure*}

\begin{figure*}
\centerline{\hbox{ \hspace{0.0in}
    \subfigure{\psfig{file=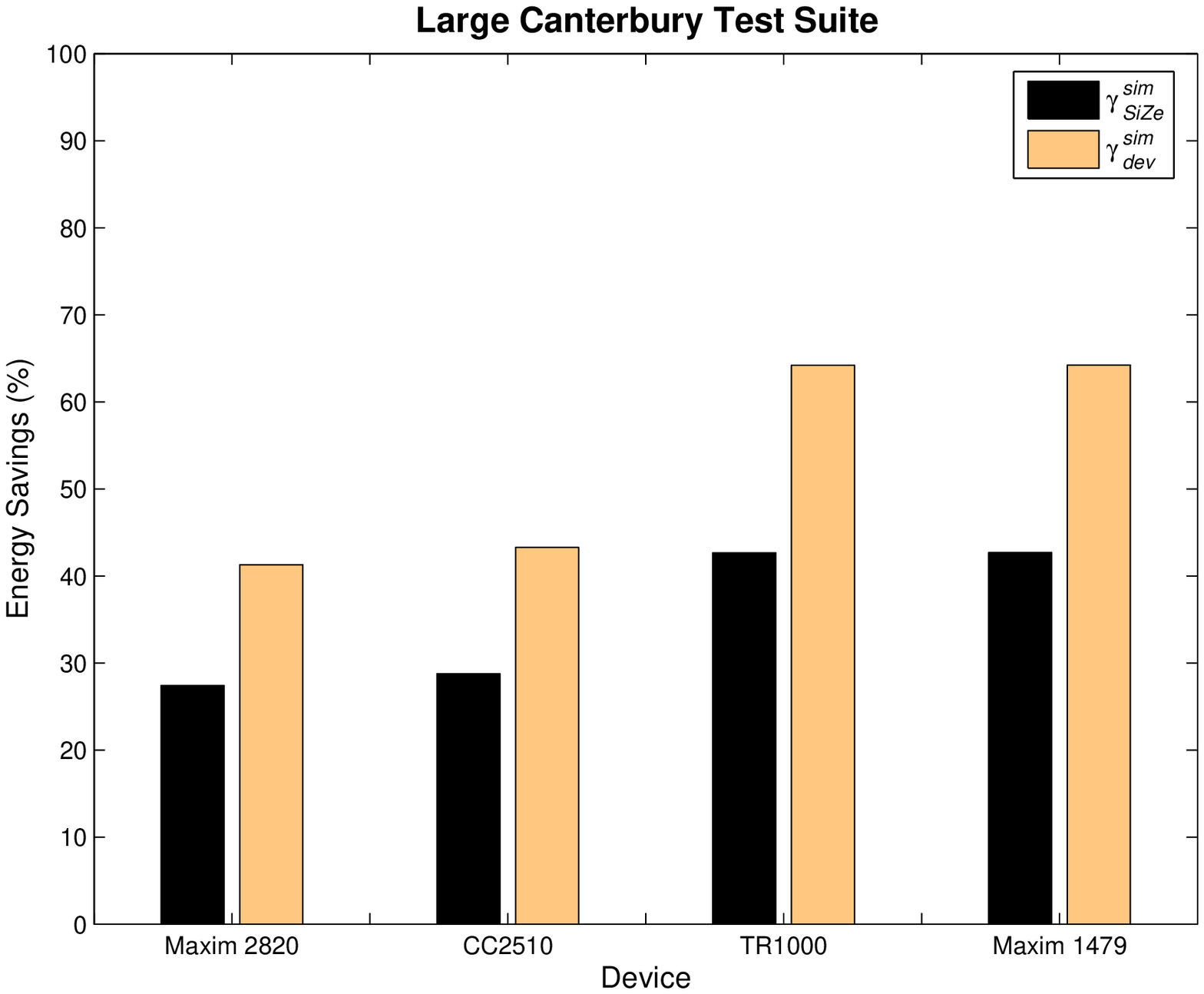,width=2.5in,height=2.5in}
    \hspace{0.13in}
    \label{figure4:a}}
    \subfigure{\psfig{file=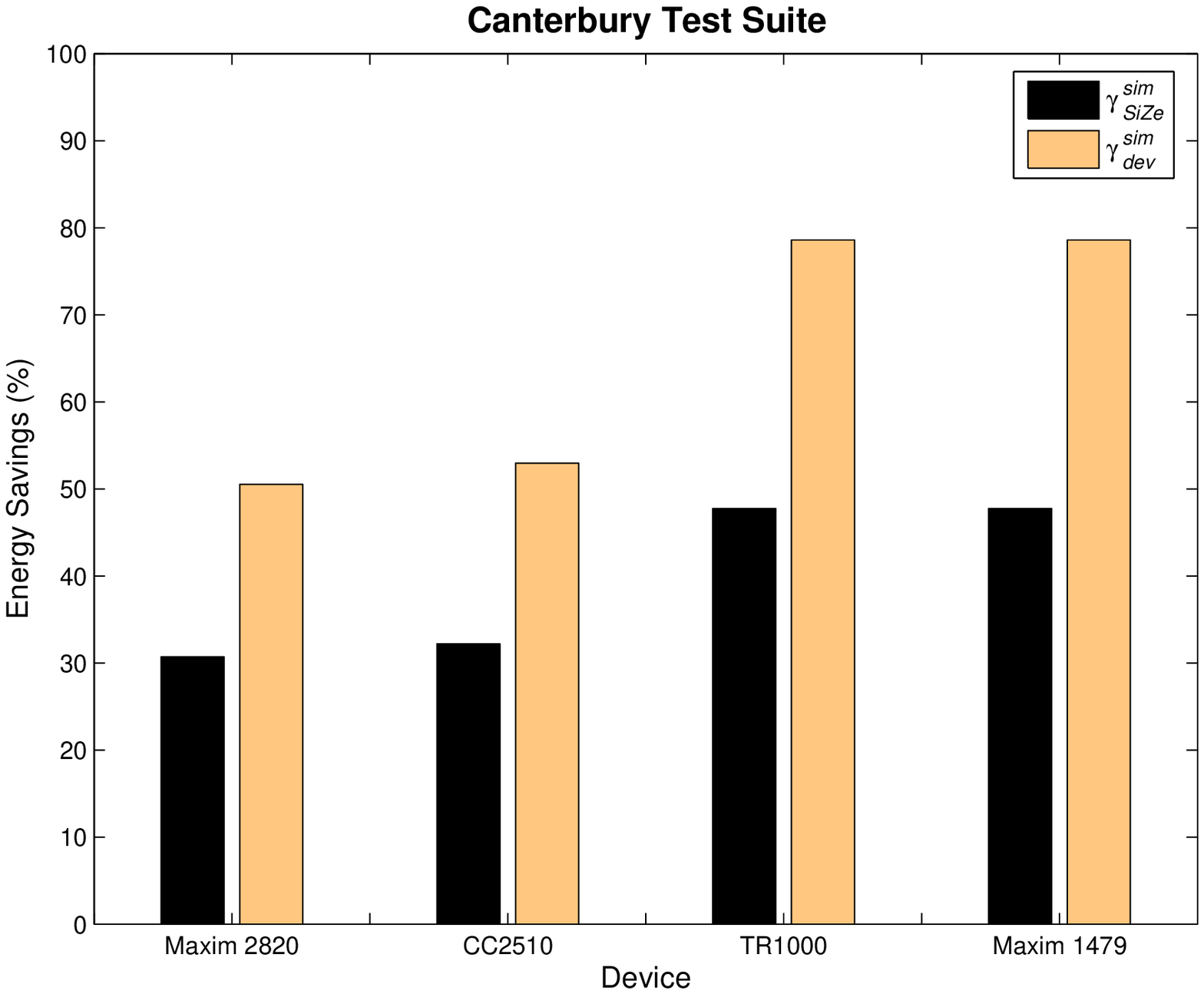,width=2.5in,height=2.5in}
    \label{figure4:b}}
    }
  }

  \hbox{\hspace{1.0in} (a) \hspace{2.67in} (b)}
  \caption{\small{Device Specific Energy Savings for the Canterbury and Large
Canterbury Test Suites}}
  \label{figure4}
\end{figure*}

The effect of real life device characteristics on the energy savings
was studied in details in \cite{sinhaaton07}. It was shown in
\cite{sinhaaton07} that the fractional energy savings generated by
the {\em TransmitRBNData} algorithm over EbT transmission scheme is,

\begin{align}
\gamma_{dev} & = \Bigl(1 - \frac{n + 2}{4n}\Bigr)\Bigl(1 -
\frac{I_{low}}{I_{high}}\Bigr) \label{gammaequation}
\end{align}

while the fractional energy savings generated by the {\em SiZe}
protocol compared to an EbT transmission scheme is given by,

\begin{align}
\gamma_{SiZe} & = \frac{I_{high} - I_{low}}{2I_{high}} = \frac{1}{2}
- \frac{I_{low}}{2I_{high}} \label{gammasize}
\end{align}

where, $I_{high}$ and $I_{low}$ denote the current drawn in the {\em
transmit} (TX) and the {\em active} states, respectively. In order
to evaluate the performance of our {\em TransmitRBNData} algorithm
on real-life devices, we considered some of the commercially
available radios for our simulation purpose. The results of our
simulation are presented in table~\ref{tabledevices}. In
table~\ref{tabledevices}, $\gamma_{SiZe}$ and $\gamma_{dev}$ refer
to the values obtained by substituting the corresponding device
parameter values in equations~\ref{gammasize} and
\ref{gammaequation}, respectively. Table~\ref{tabledevices} shows
that $\gamma_{dev}$ is higher than $\gamma_{SiZe}$ by at least 16\%
for Maxim 2820 and Chipcon CC2510Fx chips, while it is higher by
nearly 25\% for RFM TR1000 and Maxim 1479. The values of
$\gamma^{sim}_{SiZe}$ and $\gamma^{sim}_{dev}$ in
figures~\ref{figure3} and \ref{figure4} refer to the energy saving
results with the SiZe transmission scheme and our proposed algorithm
respectively, obtained by running the simulation on the benchmark
suites with the corresponding device parameters. The results show
that for the TransmitRBNData algorithm, the performance in energy
savings is always much better than the SiZe transmission scheme.
However, while the graphs of RFM TR1000 and the Maxim 1479 devices
show savings that are nearly equal to those reported in
figure~\ref{figure2}, the savings are somewhat lower for the
CC2510Fx and Maxim 2820, due to the fact that the current drawn in
the active state for both of these devices is not negligible
compared to the current in the TX state.

\section{Medium Access Control} \label{phyissues}

We present in this section a medium access control for the {\em
TransmitRBNData} algorithm. Issues related to the design of a MAC
protocol such as representation of RBN encoded numbers in the
internal buffers at the MAC layer, consideration of suitable
modulation schemes for the {\em TransmitRBNData} algorithm and
receiver-transmitter synchronization for the duration of
transmission of a data packet were addressed in \cite{sinhaaton07}.

\subsection{RBNSiZeMAC - Asynchronous MAC Protocol}

We now present RBNSiZeMAC - an asynchronous MAC protocol that allows
the transmission of RBN encoded data, for single channel wireless
networks. In order to transmit a frame successfully, a node must
compete with other nodes within its neighborhood to win the channel
for the time duration it requires to transmit its frame. It is
important to prevent simultaneous transmission to the same receiving
node as that would garble the frames beyond recovery.

\begin{figure*}
\centerline{\psfig{file=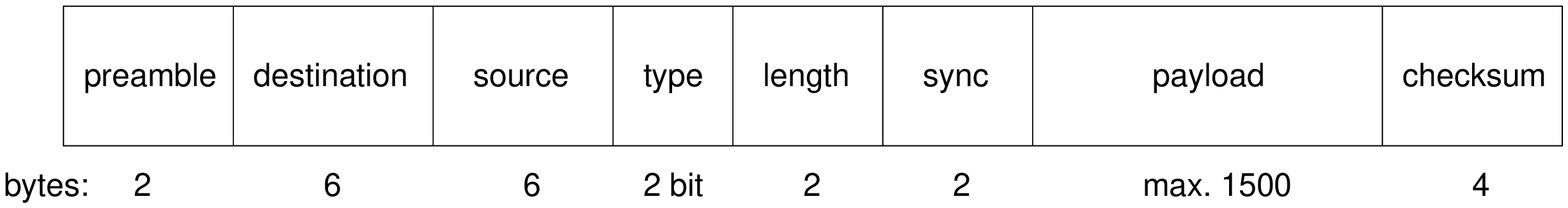,width=5in,height=0.4in}}
\centering \caption{Data frame of RBNSiZeMAC protocol}
\label{dataframe}
\end{figure*}

Our protocol uses two classes of frames : data and control. The
format of the data frame is shown in figure~\ref{dataframe}. The
{\em data frame} consists of three parts : i) a header ii) payload
and, iii) a frame trailer.  The frame header and the trailer are in
binary while the payload is in RBN. The header part consists of a
preamble, destination and source addresses, type, length and sync
fields. Each frame starts with a {\em Preamble} of 2 bytes, each
containing the bit pattern 10101010, similar to that of the IEEE
802.3 MAC header. The receiver synchronizes its clock with that of
the sender on this preamble field \cite{tannenbaum}. The next two
fields in the frame header are the {\em destination} and the {\em
source} addresses respectively, each 6 bytes long. As in the 802.3
standard, the high order bit of the destination address is a 0 for
ordinary addresses and 1 for group addresses. A frame with
destination address consisting of all 1s is accepted by all nodes in
the neighborhood of the transmitter. Next comes a 2 bit {\em type}
field which indicates whether this is a data or a control frame. The
value '00' is used to indicate a data frame while the remaining 3
possible bit patterns (01, 10 and 11) are reserved for the control
frames. The 2 byte {\em length} field gives the length of the {\em
payload} part of the frame. Finally, there are another 2 bytes of
{\em sync} field, each consisting of the binary bit pattern 10101010
to ensure that the receiver and the sender clocks remain
synchronized during the entire period of transmission of the
payload. This is necessary as we assume that no signal is
transmitted by the sender for the symbol $0$ in the RBN encoded
payload. As our proposed algorithm {\em TransmitRBNData} can lead to
the creation of long runs of 0's in the RBN encoded data, it is
important to ensure that the sender and receiver clocks remain
synchronized for the entire duration of the transmission of the
payload. The payload field can be maximum 1500 bytes, again similar
to the 802.3 frame. The payload part is followed by the binary
encoded {\em frame trailer} which consists of a 4 byte {\em
checksum} field computed only on the binary equivalent of the
payload of the frame. The checksum algorithm is the standard cyclic
redundancy check (CRC) used by IEEE 802.

\begin{figure*}
\centerline{\psfig{file=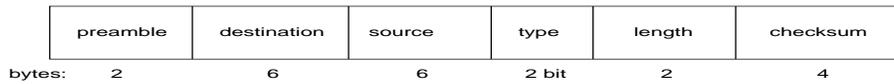,width=4.7in,height=0.4in}}
\centering \caption{Control frame of RBNSiZeMAC protocol}
\label{controlframe}
\end{figure*}

The format of the {\em control frame} is shown in
figure~\ref{controlframe}. The control frame is in binary. It
consists of a 2 byte {\em preamble}, similar to that in a data
frame, followed by the destination, source and the type fields
respectively. The size and interpretation of these three fields are
same as in the data frame. The {\em type} field is set according to
the type of the intended control message, namely RTS, CTS or ACK.
The way these three message types are used is exactly the same as in
the 802.11 MAC protocol \cite{tannenbaum}. This is followed by a
{\em length} field which indicates the length of the payload that a
node wishes to send in the data frame. In the case of a CTS or ACK
frame from a receiving node, if the receiver too has some data to
send to the sender, it sets the length field accordingly, otherwise
it is set to zero. The last field is the usual 4 byte checksum
computed over the rest of the control frame fields.

Each node in the network is assumed to possess channel status
sensing capability - i.e., whether the channel is idle or busy. The
protocol that we propose here is a CSMA/CA (CSMA with collision
avoidance) protocol, very similar to the 802.11 protocol. When a
node $A$ wants to transmit, it senses the channel. If it remains
idle for a time period of $b$, where $b$ is the maximum possible
duration of a frame transmission in RBNSiZeMAC protocol, $A$ sends
an RTS (Request to Send) control message to the receiver (say $B$).
The reason for waiting for at least $b$ time is to avoid
interrupting any ongoing frame transmission. Due to encoding of the
payload in RBN, it is possible that there may be a long run of the
symbol $0$ in the data which may otherwise be wrongly interpreted as
the channel being idle without any ongoing transmission. If $B$
receives the RTS, it may decide to grant permission to $A$ to
transmit, in which case it sends a CTS frame back. If $A$ does not
receive any CTS from $B$ or a collision occurs for the RTS frame,
each of the colliding nodes waits for a random time, using the
binary exponential back-off algorithm, and then retry. The behavior
of other nodes in the vicinity of both $A$ and $B$ on hearing the
RTS or CTS frames is the same as in 802.11.

After a frame has been sent, there is a certain amount of dead time
before any node may send a frame. We define two different intervals,
similar to the 802.11 \cite{tannenbaum}, for the RBNSiZeMAC protocol
:

\begin{enumerate}

\item{The shortest interval is SIFS (Short InterFrame Spacing) that
is used in exactly the same way as in 802.11. After a SIFS interval,
the receiver can send a CTS in response to an RTS or an ACK to
indicate a correctly received data frame.}

\item{If there is no transmission after a SIFS interval has elapsed
and a time NIFS (Normal InterFrame Spacing) elapses, any node may
attempt to acquire the channel to send a new frame in the manner
described previously. We use NIFS in exactly the same way as the
802.11 protocol uses the DIFS (DCF InterFrame Spacing) interval.}

\end{enumerate}

\section{Conclusion} \label{conclusion}

The redundant binary number system can be used instead of the binary
number system in order to increase the number of zero bits in the
data. Coupled with this, the use of silent periods for communicating
the 0's in the bit pattern provides a significant amount of energy
savings in data transmissions. The transmission time also remains
linear in the number of bits used for data representation, as in the
binary number system. Simulation results on various benchmark suites
show that with ideal as well as some commercial device
characteristics, our proposed algorithm offers a reduction in energy
consumption of 69\% on an average, when compared to existing energy
based transmission schemes. Based on this transmission strategy, we
have designed a MAC protocol that would support the communication of
such RBN encoded data frames for asynchronous communication in a
wireless network.

\end{document}